\documentclass[aps,twocolumn,showpacs]{revtex4}

\usepackage{epsfig}
\usepackage{amsfonts}
\usepackage{amssymb}
\usepackage{mathrsfs}
\usepackage{theorem}
\usepackage{amsmath}
\usepackage{times}
\usepackage{color}

\usepackage{ifpdf}
\ifpdf
\usepackage{epstopdf}   
\fi

\newtheorem{theorem}{Theorem}
\newtheorem{defin}{Definition}
\newtheorem{lem}{Lemma}
\newtheorem{corollary}{Corollary}

\newcommand{\field}[1]{\mathbb{#1}} % requires amsfonts

\newcommand{\ket}[1]{\ensuremath{\vert#1\rangle}}
\newcommand{\bra}[1]{\ensuremath{\langle #1\vert}}
\newcommand{\bk}[2]{\ensuremath{\langle #1\vert #2\rangle}}
\newcommand{\kb}[2]{\ensuremath{\vert #1 \rangle \langle #2 \vert}}

\newcommand{\tr}{\ensuremath{\mathrm{tr}}}

\def\unity{\mbox{\small 1} \!\! \mbox{1}}

\def\unity{\mbox{\small 1} \!\! \mbox{1}}

\begin{document}
\title{Optimal Entangling Capacity of Dynamical Processes}

\author{Earl T. Campbell}
\affiliation{Department of Physics and Astronomy, University College London, Gower Street, London, WC1E 6BT, UK.}
\pacs{03.67.Mn}

\begin{abstract}
We investigate the entangling capacity of dynamical operations when provided with local ancilla.  A comparison is made between the entangling capacity with and without the assistance of prior entanglement.  An analytic solution is found for the log-negativity entangling capacity of two-qubit gates, which equals the entanglement of the Choi matrix isomorphic to the unitary operator.  Surprisingly, the availability of prior entanglement does not affect this result; a property we call resource independence of the entangling capacity.  We prove several useful upper-bounds on the entangling capacity that hold for general qudit dynamical operations, and for a whole family of entanglement monotones including log-negativity and log-robustness.  The log-robustness entangling capacity is shown to be resource independent for general dynamics.   We provide numerical results supporting a conjecture that the log-negativity entangling capacity is resource independence for all two-qudit unitaries.

\end{abstract}

\maketitle  

Quantum information theory, and its various practical applications, has lead to a mature theory of entanglement as a resource.  It is now common to think of entanglement as almost a fungible commodity.  Two different quantum states with manifestly very different non-locality properties may sometimes be converted into each other by local operations, they can also be diluted~\cite{BBPS01a} and distilled~\cite{BBPSSW01a}.  The laws governing the capabilities of local operations to interchange between different entangled states are understood in terms of entanglement measures and monotones that quantify the amount of entanglement in a given quantum state~\cite{HorodeckiReview}.

Entanglement can be created between physical systems where there is a suitable non-local dynamical process.  As we can quantify the entanglement of quantum states, we can quantify the efficacy of a dynamical operation at producing entanglement.  Such quantities are called the entangling power, capacity or strength~\cite{Nielsen03} of a dynamical process.  The \textit{average} entanglement generated by unitaries, commonly called the entangling \textit{power}, has been studied by Zanardi and coworkers~\cite{Zanardi00,Zanardi01,Zanardi02}.  Whereas the \textit{maximum} entanglement that can be produced by unitaries is usually called the entangling \textit{capacity}~\cite{Kraus01,Leifer03,Chefles05}.  In the practical context of attempting to maximize entanglement production for utilization as part of a quantum information protocol, the most appropriate quantity is the entanglement capacity (herein the EC).

Kraus and Cirac considered the EC for two-qubit unitaries acting on two-qubit product states~\cite{Kraus01}.  Leifer \textit{et al} extended this analysis to find the entangling capacity when the initial two-qubit state has prior entanglement~\cite{Leifer03}, and found that this can boost the entangling capacity.  In both papers it was observed that access to local ancilla can enable much higher entangling capacities for some unitaries, with the swap gate as the most striking example.  Without local ancilla the swap gate has zero entangling capacity, but with local ancilla the swap gate can produce two Bell pairs.  Despite numerous examples where local ancilla prove beneficial, no previous work has established an analytic solution for the entangling capacity with access to local ancilla.

\begin{figure}
\includegraphics{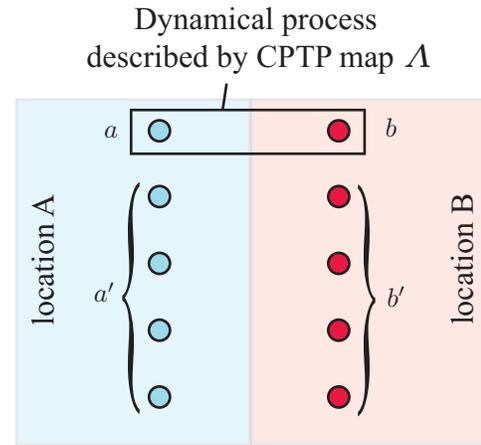}
\caption{Many architectures for quantum communication and distributed quantum computation have the above structure, possibly repeatedly over many locations.  Physical systems $A$ and $B$ cannot freely interact (e.g. because of spatial separation).  However, they can interact by a fixed dynamical process described by a completely-positive trace-preserving map, $\Lambda$, which acts on Hilbert space $H_{a} \otimes H_{b}$.  Furthermore, physical systems $A$ and $B$ also contain quantum systems  $H_{a'}$ and $H_{b'}$ respectively.  For example,  both $A$ and $B$ may contain many identical qubits, such as in two ion traps containing many atoms~\cite{Oi06}.   We show finite ancillary Hilbert space, but for simplicity assume arbitrarily large local Hilbert spaces.  Furthermore, we assume unitaries and measurements within each location can be implemented with arbitrary precision.}
\label{FIG:DistributedComputation}
\end{figure}

Furthermore, optimising the entanglement produced from a dynamical process, with access to local ancilla, is a problem that occurs in a broad range of architectures for quantum technologies~\cite{Camp2010review,Benjamin2009review,Oi06}.  Figure~\ref{FIG:DistributedComputation} shows the essential components of a physical system where this optimisation is applicable.  For example, this structure is embodied by two separate eletromagnetic traps each containing many ions, with a fixed optical process for producing entanglement between traps~\cite{Oi06}.

% Maximizing the production of entanglement, increasing the capability for executing quantum information protocols that require shared entanglement between $A$ and $B$.  

We prove that two qubit unitaries have an EC, measured by the log-negativity~\cite{Vidal02,Eisert01}, that has a simple closed form.  Interestingly this EC equals the entanglement of the \textit{Choi matrix} $\rho_{U}$ isomorphic to $U$ via the Choi-Jamiolkowski isomorphism~\cite{Choi75,Jam01a}.  This proof will also hold for a restricted class of higher dimensional unitaries that we characterize.  Cirac \textit{et al}~\cite{Cirac01} have previously proposed quantifying capabilities of entangling unitaries  by considering the entanglement of the Choi matrix.  However, Cirac  \textit{et al} did not show that the Choi matrix captures the maximum achievable entanglement for any continuous monotones.  Our results give a concrete operational meaning to the entanglement of the Choi matrix. 

\begin{figure}
\includegraphics{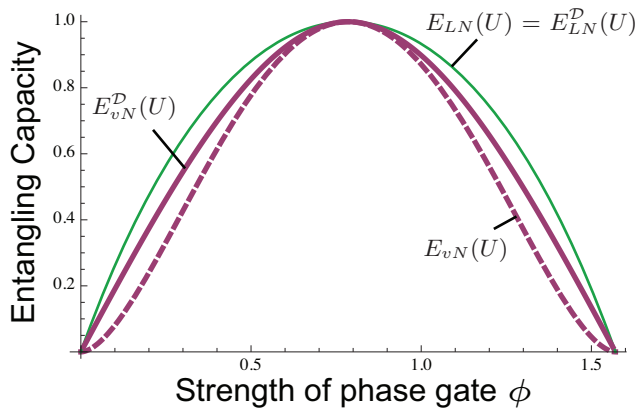}
\caption{The maximum increase in entanglement created by a phase gate, $U=\exp^{i \phi Z_{A}Z_{B}}$, as measured by: (thick dashed) the von Neumann entropy with initially separable states;  (thick solid)  the von Neumann entropy assisted by prior entanglement; (thin) the log-negativity both with and without access to prior entanglement.}
\label{fig_Phase_gate}
\end{figure}

It is surprising that our result for two-qubit unitaries holds with or without the assistance of prior entanglement.  We call this property \textit{resource independence} of the log-negativity EC for two-qubits.  In deriving our result for two-qubit unitaries, we prove two theorems that are useful in their own right. These theorems provide upper-bounds on the EC in the more general setting of two-qudit completely-positive trace-preserving maps.  The first upper-bound we prove also applies to a whole family of EC monotones that includes both log-negativity EC and log-robustness EC.  Our upper-bounds entail that log-robustness EC is always resource independent, and give us reason to suspect that log-negativity EC is also generally resource independent.  We support this conjecture with a numerical study of the log-negativity EC for two-qutrit unitaries, where we found no violations of resource independence.

\section{Notation and Definitions}

We aim to solve the following problem:  for a given completely-positive trace-preserving (CPTP) map $\Lambda$ acting on Hilbert spaces $H_{a}\otimes H_{b}$, what is the maximum entanglement that can be produced?  This maximization problem depends on what resources are available.   Throughout, we assume that local ancilla (typically labelled as Hilbert space $H_{a'}\otimes H_{b'}$), local unitaries and classical communication are freely available as is appropriate in many communication and distributed computation scenarios.     The maximum achievable entanglement for a dynamical map, $\Lambda$, when the initial state is separable, is given by:
\begin{equation}  
	E( \Lambda ) = \sup \{   E(  \Lambda ( \sigma ) )    | \sigma \in \mathcal{S}  \} ,
\end{equation}
where $\mathcal{S}$ denotes the set of separable density matrices, and $E(\rho)$ is an appropriate entanglement monotone for density matrices.  However, when prior entanglement is available, the  maximum increase of entanglement is:
\begin{equation}  
	E^{\mathcal{D}}( \Lambda ) = \sup \{   E(  \Lambda ( \sigma ) )  - E (\sigma)  | \sigma \in \mathcal{D}  \} ,
\end{equation}
where $\mathcal{D}$ is the set of all physical density matrices, and we have denoted the availability of this resource by a superscript. This quantity cannot be smaller than the unassisted capacity, $E(\Lambda)$, but in general may be larger. The von Neumann entropy~\footnote{When using the von Neumann entropy one must assume only pure states and unitary gates.}, $E_{vN}$, is an example where some unitary gates have been shown to exhibit an increase, $E_{vN}^{\mathcal{D}}(U)>E_{vN}(U)$, when assisted by entanglement~\cite{Dur01}.  We do not need to invoke exotic unitary gates to observe the phenomena, as it occurs even for simple two-qubit phase gates, as shown in Fig.~\ref{fig_Phase_gate}.  Alternatively, we may be interested in knowing how much entanglement could be produced given access to a restricted class of resources, such as bound entanglement~\cite{HoroBound}:
\begin{equation}
	E^{\mathcal{B}}( \Lambda ) = \sup \{   E(  \Lambda ( \sigma ) )  - E (\sigma)  | \sigma \in \mathcal{B}  \} ,
\end{equation}
where $\mathcal{B}$ is the set of all physical density matrices with only bound entanglement.  Formally, all operators in $\mathcal{B}$ are positive semi-definite before and after partial transposition.

Since the von Neumann EC violates resource independence, such that $E_{vN}^{\mathcal{D}} \neq E_{vN} $, there is a temptation to infer that this phenomena would be observed for other EC monotones.  Here we present results that counter this intuition, showing that many other entanglement monotones have an inherent resource independence, which makes them cleaner, and simpler, candidates for EC monotones.

We are principally interested in the logarithmic versions of negativity~\cite{Vidal02,Eisert01} and robustness~\cite{Vidal99} and other similar decomposition-based monotones (defined in section \ref{sec_norm_based_measures}.).  The negativity is defined as:
\begin{equation}
\label{eqn:neg}
	E_{N} ( \rho ) = ( || \rho^{\Gamma} || - 1)/2
\end{equation}
where $\Gamma$ denotes a partial transpose over system $B$, and $||...||$ denotes the trace norm:
\begin{equation}
	|| A || = \tr \left(  \sqrt{ A^{\dagger} A } \right).
\end{equation}
The robustness of $\rho$ was first defined~\cite{Vidal99} as the amount of any separable state $\sigma'$ that must be mixed with $\rho$ to make the whole mixture separable:
\begin{equation}
	E_{S} ( \rho) = \inf \{ t | \sigma = \frac{\rho + t \sigma' }{1+ t} ; \sigma, \sigma' \in \mathcal{S} \} ,
\end{equation}
where $\mathcal{S}$ is again the set of separable states.  Despite the contrast between the conventional definitions of $E_{N}$ and $E_{S}$, we will see that they are actually very closely related~\cite{Vidal02}.  

Throughout, we are interested in logarithmically rescaled variants of both monotones as $E_{Lx}(\rho)=\log_{2}(1+2E_{x}(\rho))$.  Logarithmic variants of negativity and robustness are strongly subadditive under tensor products, such that:
\begin{equation}
	E ( \rho_{1} \otimes \rho_{2} ) \leq E (\rho_{1}) + E (\rho_{2}). 
\end{equation}
We desire this property as any monotone violating this condition can never give rise to an EC that is resource independent (see Appendix~\ref{APP:subadd_necessary}).  Indeed, the concept of resource assisted EC can become meaningless when subadditivity is not respected.  Conversely, if an EC is resource independent, then a rescaling of the underlying entanglement monotone that destroys subadditivity must also remove resource independence.  Under fairly weak assumptions such a rescaling can always be found (see Appendix~\ref{APP:always_break_subadd}).  

When a subadditive entanglement monotone always saturates the inequality, such that
\begin{equation}
	E_{LN}(  \rho_{1} \otimes \rho_{2} ) =	E_{LN}(  \rho_{1}  ) + E_{LN}(  \rho_{2}  ) ,
\end{equation}
the entanglement monotone is said to be strongly additive.    Strong additivity gives entanglement monotones a clearer operational interpretation, and so is typically desirable.   Any EC monotone that is both resource independent and strongly additive has the following elegant property: given a dynamical map $\Lambda$ with entangling capacity, $E(\Lambda)$, we can use the dynamical map $n$ times to produce, at best, exactly $n E(\Lambda)$ entanglement.  The log-negativity is both strongly additive and provides an upperbound on the number of singlets that can be distilled~\cite{Horodecki00}, and so we will focus on this monotone.

Note that, our notation for EC monotones always uses a calligraphic superscript to denote the resources that are relevant to the maximization problem, and a subscript to describe the underlying entanglement monotone for quantum states.  

\section{Decomposition-based monotones}
\label{sec_norm_based_measures}

\begin{figure}
\includegraphics{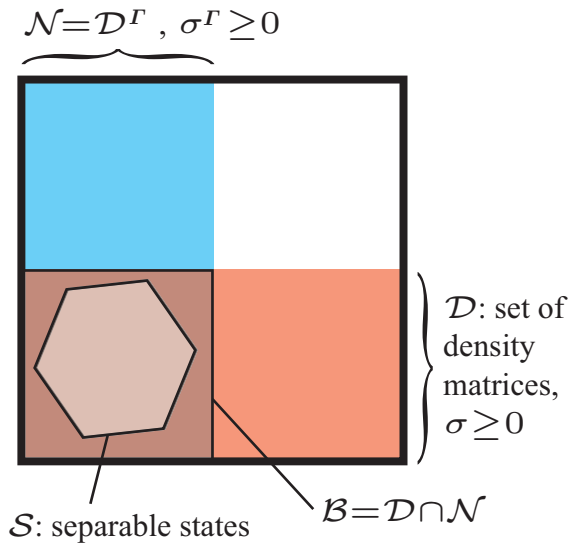}
\caption{An heuristic diagram of different sets of zero-entanglement operators:  operators in $\mathcal{D}$ have non-negative eigenvalues; operators in $\mathcal{N}$ have partial transposes with non-negative eigenvalues; operators in $\mathcal{B}$ have both the previous properties; operators in $\mathcal{S}$ can be decomposed as a positive sum of separable operators.  The whole space shown represents all Hermitian unit-trace matrices.}
\label{fig_nestings}
\end{figure}

Here we introduce the idea of decomposition-based monotones of entanglement, as first discovered by Werner and Vidal~\cite{Vidal02}.    They proposed entanglement monotones of the form:
\begin{equation}
\label{EQN:decomp-based}
	E_{\mathcal{M}}(  \rho  )   = \inf \{  t \vert \rho = (1+t ) \sigma^{+} - t \sigma^{-} ; t \geq 0 ; \sigma^{\pm} \in \mathcal{M}   \} .
\end{equation}
This formula returns the minimum value of $t$ over all real linear decompositions of $\rho$ into matrices $\sigma^{\pm}$ that belong to a specific set $\mathcal{M}$.  We will call operators in the set $\mathcal{M}$ zero-entanglement operators as for all $\sigma \in \mathcal{M}$ we have $E_{\mathcal{M}}(  \sigma   )=0 $.   However, remember that a zero-entanglement operator is not necessarily separable, as the monotone may fail to detect entanglement for some states.  In order to ensure that the quantity is well-defined and monotonic under local operations, we require that the set of zero-entanglement operators has the following properties~\footnote{We have slightly amended the list of properties to ensure that $E_{\mathcal{M}}(\rho_{A} \otimes \rho_{B} )=0$ and $E_{\mathcal{M}}\left( (U_{A} \otimes U_{B}) \rho (U_{A} \otimes U_{B})^{\dagger} \right)=E_{\mathcal{M}}( \rho )$, which did not necessarily follow from the original formulation.}:
\begin{enumerate}
	\item the set $\mathcal{M}$ is a compact and convex, such that $p \sigma + (1-p)\sigma' \in \mathcal{M}$, whenever $\sigma, \sigma' \in \mathcal{M}$ and $0 \leq p \leq 1$;
	\item all matrices in $\mathcal{M}$ are Hermitian and unit-trace, so $\sigma=\sigma^{\dagger}$ and $\tr(\sigma)=1$;
	\item the set $\mathcal{M}$ includes all separable states, $\mathcal{S} \subseteq \mathcal{M}$;
	\item the set is invariant under local unitaries, $(U_{A} \otimes U_{B})\mathcal{M}(U_{A}\otimes U_{B})^{\dagger}=\mathcal{M}$.
\end{enumerate}
When these conditions hold, any Hermitian unit-trace matrix, $\rho$, can always be decomposed into zero-entanglement operators,  $\rho=(1+t ) \sigma^{+} - t \sigma^{-} $.  This result is well known for the set of separable states~\cite{Vidal99}, $\mathcal{S}$,  and so will also hold for any set that contains $\mathcal{S}$.  Collectively, these properties ensure that $E_{\mathcal{M}}$ is always an entanglement monotone, and similarly~\footnote{After taking the logarithm, the monotone is no longer convex.  However, Plenio~\cite{Plenio05} showed that the log-negativity is a full entanglement monotone despite lacking convexity.  His argument uses the generic properties of $E_{N}=E_{\mathcal{N}}$ that also hold for our generalized family entanglement monotones, and so the argument carries over.} for logarithmic quantities, which we denote:
\begin{equation}
\label{eqn:Log}
E_{L\mathcal{M}}(  \rho ) = \log_{2} ( 1+2 	E_{\mathcal{M}}(  \rho  )  ) .
\end{equation}
Familiar monotones in this family include negativity and robustness of entanglement, which are:
\begin{eqnarray*}
	E_{S} ( \rho ) & = & 	E_{\mathcal{S}}(  \rho  ) , \\ \nonumber
	E_{N} ( \rho ) & = & 	E_{\mathcal{N}}(  \rho  ) ,
\end{eqnarray*}
where $\mathcal{S}$ is the set of separable matrices, and $\mathcal{N}=\mathcal{D}^{\Gamma}$.  The equivalence for robustness is immediate, but for negativity requires a little work.  The more familiar definition of negativity is in terms of the trace norm and partial transpose.  For any unit-trace hermitian operator, there is always decomposition into positive unit-trace operators $\rho=(1+t) \sigma^{+} - t \sigma^{-}$.  We can always find such a decomposition where $\sigma^{\pm}$ are orthogonal, and then the trace-norm is simply $2t+1$.  Clearly no smaller value can be achieved, and so finding the trace norm is equilvalent to a minimization problem:
\begin{equation}
	|| \rho || = \inf \{ 2t+1 \vert \rho = t \sigma^{+} - (1-t) \sigma^{-} ; \sigma^{\pm} \in \mathcal{D} \}
\end{equation}
However, when calculating the negativity we take the partial transpose first:
\begin{eqnarray}
	|| \rho^{\Gamma} || & = & \inf \{ 2t+1 \vert \rho^{\Gamma} = (1+t) \sigma^{+} - t \sigma^{-} ; \sigma^{\pm} \in \mathcal{D} \} , \\ \nonumber
	 & = & \inf \{ 2t+1 \vert \rho = t \sigma^{+} - (1-t) \sigma^{-} ; \sigma^{\pm} \in \mathcal{D}^{\Gamma}=\mathcal{N} \} .
\end{eqnarray}
Hence, when calculating the negativity, instead of taking the partial transpose we can find the minimal decomposition w.r.t the partial transpose of the physical density operators.  

In addition to more familiar entanglement monotones, Vidal and Werner also discussed an intermediate monotone for the set $\mathcal{B}$ that contains $\sigma$ that are positive and have a positive partial transpose.  They observed that these monotones are related by the inequalities:
\begin{equation}
E_{\mathcal{S}}(  \rho  ) \geq 	E_{\mathcal{B}}(  \rho  ) \geq 	E_{\mathcal{N}}(  \rho  )  \geq 0 ,
\end{equation}
which follows from the inclusions $\mathcal{S} \subset \mathcal{B} \subset \mathcal{N}$.  For an overview of how these sets relate see figure.~\ref{fig_nestings}.

\section{The duality theorem}
\label{sec_limits}

Having introduced the family of decomposition-based entanglement monotones, we now formulate our first upper-bound on the corresponding EC monotones.
\begin{theorem}
\label{infMeasures}
For any logarithmic decomposition-based entanglement monotone, $E_{L\mathcal{M}}(\rho)$, the corresponding entangling capacity of a CPTP map $\Lambda$ satisfies:
\begin{equation}
	E^{\mathcal{D}}_{L\mathcal{M}}( \Lambda ) \leq  E_{L \mathcal{M}}^{\mathcal{M}}(  \Lambda ) = \sup \{ E_{L \mathcal{M}} ( \Lambda( \sigma ) ) | \sigma \in \mathcal{M} \}.
\end{equation}
Furthermore if $\mathcal{M} \subseteq \mathcal{D}$, where $\mathcal{D}$ is the set of physical density matrices, then we have equality, and hence:
\begin{equation}
	E^{\mathcal{D}}_{L \mathcal{M}}( \Lambda ) =  E_{L \mathcal{M}}^{\mathcal{M}}(  \Lambda ) .
\end{equation}
\end{theorem}
This theorem instructs us that the resource assisted EC can never exceed the entanglement produced from applying the CPTP map to an operator with zero-entanglement w.r.t. the entanglement monotone used.  

Consequently, a rough interpretation of this theorem is that the EC monotones considered are inherently resource independent.  However, this is only true when the above equality holds, as it does for log-robustness EC.  However, here only the inequality is proven to apply for log-negativity EC.  This is because the set of zero-entanglement operators, $\mathcal{N}$, includes unphysical operators with negative eigenvalues (formally $\mathcal{N} \not \subseteq \mathcal{D}$).  Clearly, when maximization over some unphysical operators is required to rigorously derive an upperbound, we do not guarantee that the upperbound is attainable.

Since the resource independence interpretation is not always strictly accurate, the more cautious reader may prefer to think of the theorem as simply relating two optimization problems.  The optimization giving the upper bound we will call, in absence of more appropriate terminology, the dual problem.  

To prove the theorem, we begin by recalling our definition of resource assisted entangling capacity:
\begin{equation}
	E^{\mathcal{D}}_{L \mathcal{M}}( \Lambda ) = \sup \{  E_{L\mathcal{M}}(  \Lambda (\rho) ) - E_{L\mathcal{M}}(\rho) | \rho \in \mathcal{D}   \} .
\end{equation}
We shall use $\rho^{*}$ to label a physical state that achieves the supremum.  The initial entanglement of this state is $E_{L\mathcal{M}}(\rho^{*}) =\log_{2} (1+2t)$, where there will exist an optimal decomposition:
\begin{equation}
\label{EQN:opt_rho_decomp}
	\rho^{*} = (1+t) \sigma ^{+} - t \sigma ^{-} ; \sigma ^{\pm} \in \mathcal{M} .
\end{equation}
Think of this as a decomposition into zero-entanglement operators.  Acting with the CPTP map gives:
\begin{eqnarray}
\label{EQN:Lambda_rho_decomp}
	\Lambda (\rho^{*}) = (1+t) \Lambda (\sigma^{+}) - t \Lambda (\sigma^{-}) ,
\end{eqnarray}
Even though $\Lambda ( \sigma^{\pm} )$ are not always physical density matrices, we can still apply the formula for calculating their entanglement and we can compare it with the entanglement produced by acting on the physical state $\rho^{*}$ .  Performing this calculation, and some algebraic manipulation (see Appendix~\ref{APP:duality_details}), we find:
\begin{equation}
\label{EQN:duality_half_step}
	E_{L \mathcal{M}}(\Lambda(\rho^{*}) ) -E_{L \mathcal{M}}( \rho^{*} )  \leq  \max \{  E_{L \mathcal{M}}(\Lambda( \sigma ^{\pm}) )  \} .
\end{equation}
This tells us that if we knew the optimal physical input, $\rho^{*}$, then the entanglement produced would not exceed the entanglement produced from the zero-entanglement operators, $\sigma^{+}$ and $\sigma^{-}$.   However,  we don't yet know what $\rho^{*}$ is!  Despite this, we can still give an upperbound by maximizing over all zero-entanglement operators, $\sigma \in \mathcal{M}$, such that:
\begin{equation}
\label{dualityproved}
	E^{\mathcal{D}}_{L \mathcal{M}}( \Lambda ) \leq \sup \{ E_{L \mathcal{M}} ( \Lambda( \sigma ) ) | \sigma \in \mathcal{M} \} ,
\end{equation}
which proves the first part of our theorem.  If all zero-entanglement operators are also physical density matrices, then the maximum is physically attainable, and the inequality is saturated.  Formally, if $\mathcal{M} \subseteq \mathcal{D}$ we have an equality, which is an especially strong result for log-robustness:
\begin{corollary}
 	The log-robustness entangling capacity of any CPTP map $\Lambda$ satisfies:
	\begin{eqnarray}
		E^{\mathcal{D}}_{L \mathcal{S}}( \Lambda ) & = & E_{L \mathcal{S}}(\Lambda)
	\end{eqnarray}
	and so the maximum possible increase is achievable with an initially separable state.
\end{corollary}

For log-negativity the result is weaker.  It tells us that the resource-assisted EC is no greater than the dual optimization of the maximum increase when the initial matrix is a, potentially unphysical, zero-entanglement operator, $\sigma \in \mathcal{N}$.   The sets $\mathcal{D}$ and $\mathcal{N}$ are two very closely related species of matrices ( note that $\mathcal{N}=\mathcal{D}^{\Gamma}$ ) and so it isn't immediately clear that we have gained much by exchanging a maximization over $\sigma \in \mathcal{D}$ for a maximization over  $\sigma \in \mathcal{N}$.  However, the duality theorem will prove its worth by playing a pivotal role in deriving the entangling capacity for two-qubit unitaries.  

From a numerical perspective, the upper bound is actually easier to work with for reasons we explain later.  For now we simply remark that numerical simulations indicate that little is lost from the relaxations needed to derive this bound, as numerical studies have not revealed any instances where the bound cannot be saturated.  We discuss this further is section~\ref{sec:Numerical}.

\section{An upper bound on log-negativity entangling capacity}
\label{sec_log_neg_upper_bound}

So far we have derived a general upper bound for a whole family of EC monotones.  Herein we focus solely on log-negativity EC.  Also, whereas the previous bound involved a maximization procedure that typically must be performed numerically, here we will derive a bound that can be evaluated by standard algebraic techniques.  The bound we find here takes a very simple form for two-qubit unitaries, with later sections demonstrating that the bound is saturated by preparing the appropriate Choi matrix.

We begin by defining some notation. For a CPTP map $\Lambda$ there will always exist a Kraus decomposition:
\begin{equation}
	\Lambda( \rho ) = \sum_{i} K_{i} \rho K_{i}^{\dagger},
\end{equation}
and each Kraus operator has a Schmidt decomposition:
\begin{equation}
	K_{i}  =  \sum_{j=1}^{d^{2}} \lambda_{j,i} A_{j, i} \otimes B_{j,i}.
\end{equation}
Recall that operator Schmidt decompositions~\footnote{Schmidt decompositions are typically defined for vectors, but the space of complex matrices is itself a vector space.  Therefore, an operator decomposition can be found by reshaping the matrix into a vector, Schmidt decomposing it in vector form, and then returning to matrix form. At least one Schmidt decomposition of an operator always exists.} always have (\textit{i})  $d=\mathrm{min}(d_{a},d_{b})$; (\textit{ii}) coefficients $\lambda$ that are positive real numbers;  and (\textit{iii}) sets of operators $\{A_{1,i},.. A_{d^{2},i} \} $ and $\{B_{1,i},... B_{2^{d},i } \}$ that are orthonormal sets, with orthonormality defined w.r.t the inner product $\langle M , N \rangle= \tr ( M^{\dagger} N )$.  Our theorem will also make use of the operator norm:
\begin{equation}
	|| M ||_{\mathrm{op}} = \sup \left\{ \frac{ \bra{\psi}  M \ket{\psi} }{ \bk{\psi}{\psi} } \right\} ,
\end{equation}
which for Hermitian $M$ is simply the largest eigenvalue.  We can now state the main result of this section:
\begin{theorem}
\label{eqn:core_result}
The log-negativity entangling capacity of any CPTP map $\Lambda$ is upper bounded by:
\begin{equation}
	E_{LN}^{\mathcal{D}}(\Lambda) \leq  \log_{2} \left(  \sum_{i} ||  O_{A, i}  ||_{\mathrm{op}} . || O_{B, i} ||_{\mathrm{op}}  \right),
\end{equation}
where:
\begin{eqnarray*}
	O_{A, i} & = & \sum_{j} \lambda_{j,i} A_{j,i}^{\dagger} A_{j,i} \\
	O_{B, i} & = & \sum_{j} \lambda_{j,i} B_{j,i}^{\dagger} B_{j,i} 
\end{eqnarray*}
\end{theorem}
For unitary maps, the CPTP map has a simple form with a single Kraus operator. The theorem simplifies even further for a particular class of unitaries that we call \textit{basic} unitaries, which we define as:
\begin{defin}
A unitary $U$ acting on $C^{d_{a}} \otimes C^{d_{b}}$ is said to be \textit{basic} if there exists an operator Schmidt decomposition:
\begin{equation}
	U = \sum_{k=1}^{d^{2}} \lambda_{k} A_{k} \otimes B_{k},
\end{equation}
where $A_{k}$ and $B_{k}$ are proportional to unitaries, such that $A_{k}^{\dagger}A_{k}= \unity / d_{a}$ and $B_{k}^{\dagger}B_{k}= \unity / d_{b}$.  
\end{defin}
For basic unitaries the related operator norms are simply $|| O_{A} ||_{\mathrm{op}}=|| O_{B} ||_{\mathrm{op}} = \sum_{i}\lambda_{i}/\sqrt{d_{a}d_{b}}$, and so our theorem tells us that:
\begin{equation}
\label{eqn:basic_unitary_bound}
	E_{LN}^{\mathcal{D}}( U ) \leq 2  \log_{2} \left(  \sum_{i} \frac{\lambda_{i}}{\sqrt{d_{a}d_{b}}}   \right).
\end{equation}
Not all unitaries are basic, but many are.  Crucially, \textit{all} two-qubit unitaries are basic  (see Appendix.~\ref{APP:basic_unitaries}), and so this result applies to many physical systems.

Having laid out the results of this section, we now turn to proving them by utilizing the duality theorem, Thm.~(\ref{infMeasures}).  We will find an upper bound on the resource-assisted EC, by considering the dual problem of maximizing the entanglement producible from zero-entanglement operators, $\sigma \in \mathcal{N}$.  Since the log-negativity is a monotonic function of the negativity,  the maximum is achieved for the same $\sigma$, so we are interested in finding the maximum of:
\begin{eqnarray}
 	|| \Lambda (\sigma)^{\Gamma} || & = & || \sum_{i} (K_{i} \sigma K_{i})^{\Gamma}  || , \\
	& \leq & \sum_{i} ||  (K_{i} \sigma K_{i})^{\Gamma}  || ,
\end{eqnarray}
with the second line following by convexity of the trace norm.  Our proof will proceed by focusing on the maximum for individual Kraus operators. For brevity, herein we drop the $i$ subscript, s.t. $K_{i} \rightarrow K$.  For these single terms we have that: 
\begin{equation} 
	K \sigma K^{\dagger} =  \sum_{k,j} \lambda_{k} \lambda_{j} (A_{k} \otimes B_{k }) \sigma ( A_{j} \otimes  B_{j})^{\dagger} ,
\end{equation}
taking the partial transpose we arrive at:
\begin{eqnarray} 
\label{EQN:Log_neg_partial_trace}
	(K \sigma K^{\dagger})^{\Gamma} & = &  \sum_{k,j} \lambda_{k} \lambda_{j} (A_{k}  \otimes B_{j}^{T}) \sigma^{\Gamma}  (A_{j} \otimes  B_{k}^{T})^{\dagger} , 
\end{eqnarray}
The next step is to find the trace norm of this expression.  As outlined in section~\ref{sec_norm_based_measures}, taking the trace norm is equivalent to a minimization problem over all decompositions into positive unit-trace operators.  By finding a decomposition, which does not necessarily give the minimum, we are able to deduce an upper bound on the trace norm.  Here, we are always able to find a decomposition (see Appendix~\ref{APP:Log_Neg_Bound}) such that the trace norm is bounded as follows:
\begin{eqnarray} 
\label{EQN:Log_neg_trace_norm}
	||(K \sigma K^{\dagger})^{\Gamma}|| & \leq &   \sum_{j, k}  \lambda_{k}  \lambda_{j} \tr ( A_{j}^{\dagger}A_{j} \otimes B_{k}^{*}B_{k}^{T} \sigma^{\Gamma} ) .
\end{eqnarray}
Making use of the more compact notation introduced earlier, this is equivalent to:
\begin{eqnarray*} 
	||(K \sigma K^{\dagger})^{\Gamma}||  & = &   \tr ( O_{A} \otimes O_{B}^{T} \sigma^{\Gamma} ) .
\end{eqnarray*}
Maximizing over all $\sigma \in \mathcal{N}$ (or $\sigma^{\Gamma} \in \mathcal{D}$ ) is equivalent to finding the operator norm, and so:
\begin{eqnarray}
\sup_{\sigma \in \mathcal{N}} ||(K \sigma K^{\dagger})^{\Gamma}|| & \leq & || O_{A} \otimes O_{B}^{T} ||_{\mathrm{op}} ,  \\ \nonumber
& = &  || O_{A} ||_{\mathrm{op}} . || O_{B}^{T} ||_{\mathrm{op}} ,  \\ \nonumber
& = &  || O_{A} ||_{\mathrm{op}} . || O_{B} ||_{\mathrm{op}} .
\end{eqnarray}
The second line follows because the operator norm is multiplicative under tensor products, and the last line follows because transposition does not change the eigenvalues of Hermitian matrices.  Taking the logarithm and applying the duality theorem (Thm.~\ref{infMeasures}), proves theorem~\ref{eqn:core_result}.  

It is important that this result holds when local ancilla are available, and we will clarify this now.  For brevity, we have denoted the Kraus operator by $K$.  However, we implied the availability of local ancilla, which is equivalent to finding the EC for an operator $K \otimes \unity_{a'} \otimes \unity_{b'}$.  For this extended Kraus operator, the appropriate bound is determined by $  || O_{A} \otimes \unity_{a'} ||_{\mathrm{op}}$ and $  || O_{B} \otimes \unity_{b'} ||_{\mathrm{op}}$.  However,  $  || O_{A} \otimes \unity_{a'} ||_{\mathrm{op}} = || O_{A} ||_{\mathrm{op}}$, and so the upper-bound does indeed allow for local ancilla.

As we remarked earlier, this result takes an especially simple form for basic unitaries, and we shall see that basic unitaries saturate the bound.   However, in section~\ref{examples} we give some explicit examples that fail to saturate it. 

\section{Choi Matrices}

\begin{figure}
\includegraphics{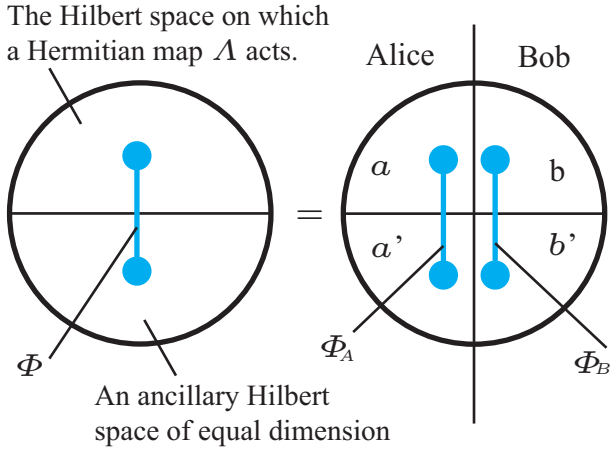}
\caption{The Hilbert spaces used in defining the Choi matrix $\rho_{\Lambda}=(\Lambda \otimes \unity)(\Phi)$ for a dynamical map, $\Lambda$.  For any bipartition,  $H_{a}\otimes H_{b}$, of the space on which $\Lambda$ acts,  we find the Choi matrix can be defined with the initial state as a separable state, $\ket{\Phi}=\ket{\Phi_{A}}\ket{\Phi_{B}}$, w.r.t this bipartition.}
\label{fig_Choi}
\end{figure}

Before showing that the bound can be saturated, we will review the concept of a Choi matrix.  The Choi-Jamiolkowski isomorphism~\cite{Choi75,Jam01a} shows that there exists a one-to-one correspondence, an isomorphism, between Hermitian matrices and Hermitian maps\footnote{A Hermitian map is a real linear sum of CPTP maps, possibly including negative contributions.}.  In particular, for every map $\Lambda$ that acts on a Hilbert space, $\field{C}^{d_{a}} \otimes \field{C}^{d_{b}}$, there exists a Hermitian matrix $\rho_{\Lambda}$ that acts on   $\field{C}^{d_{a}} \otimes \field{C}^{d_{b}}$. We call $\rho_{\Lambda}$ the Choi matrix for $\Lambda$, and it is defined by:
\begin{equation}
	\rho_{\Lambda}  =  ( \Lambda \otimes \unity )( \Phi ) ,
\end{equation}
where $\Phi=\kb{\Phi}{\Phi}$, and:
\begin{equation}
	\ket{\Phi}=\sum_{x=1}^{d_{a} d_{b} } \ket{x}_{a,b} \ket{x}_{a', b'}  /  \sqrt{d_{a}d_{b}} .
\end{equation}
This state is entangled between Hilbert spaces $H_{a, b}$ and $H_{a', b'}$.  However, the initial state, $\Phi$,  can be separable with respective to a partition between $H_{a, a'}$ and $H_{b, b'}$, by choosing the basis such that $\ket{x}_{a,b}=\ket{j}_{a}\ket{k}_{b}$ and  $\ket{x}_{a',b'}=\ket{j}_{a'}\ket{k}_{b'}$ (see figure~\ref{fig_Choi}).   It then follows that:
\begin{equation}
	\ket{\Phi}=\sum_{j=1}^{d_{a}} \sum_{k=1}^{d_{b}} \ket{j}_{a} \ket{k}_{b} \ket{j}_{a'} \ket{k}_{b'}  /  \sqrt{d_{a}d_{b}} .
\end{equation}
Switching the order of the qubits from $a,b,a',b'$ to $a,a',b,b'$, we arrive at:
\begin{eqnarray}
	\ket{\Phi} & = & \sum_{j=1}^{d_{a}}\sum_{k=1}^{d_{b}} \ket{j}_{a} \ket{j}_{a'} \ket{k}_{b} \ket{k}_{b'}  / \sqrt{d_{a} d_{b}} , \\ \nonumber
		& = & \left( \sum_{j=1}^{d_{a}} \frac{\ket{j}_{a} \ket{j}_{a'}}{\sqrt{d_{a}}} \right) \left( \sum_{k=1}^{d_{b}} \frac{\ket{k}_{b} \ket{k}_{b'}}{\sqrt{d_{b}}} \right) ,
\end{eqnarray}
which is simply a separable state:
\begin{equation}
	\ket{\Phi}= \ket{\Phi_{A}}\ket{\Phi_{B}}.
\end{equation}
As such the Choi matrix is both mathematically intriguing \textit{and} operationally meaningful as it can be prepared from a separable state and a single use of a dynamical map. 

\section{Choi Matrix Entanglement}
\label{sec_choi_ent}

Now we show that basic unitaries saturate the upper-bound described by theorem~(\ref{eqn:core_result}), and more concisely by equation~(\ref{eqn:basic_unitary_bound}).  We achieve this by using a basic unitary to reproduce its Choi matrix.  Operationally, the Choi matrix is produced by applying $U$ to the initial separable state, $\ket{\Phi_{A}} \ket{\Phi_{B}}$.  Establishing the entanglement of the Choi matrix doesn't actually rely on the operator being a basic unitary, and can be considered as a useful lower bound for the EC of any general Kraus operator.  Specifically, we show that:
\begin{theorem}
\label{thm:Choi_state}
The log-negativity entangling capacity of any Kraus operator $K$ satisfies:
\begin{equation}
	 E_{LN}^{\mathcal{D}}( K ) \geq  E_{LN}( K ) \geq 2  \log_{2} \left( \sum_{k} \frac{ \lambda_{k} }{\sqrt{d_{a}d_{b}}} \right) ,	
\end{equation}
where $\lambda_{j}$ are the Schmidt coefficients of $K$.   Equality holds if $K$ is a basic unitary, such as a two-qubit unitary.
\end{theorem}
It is important to remember that we can produce the Choi state from an initially separable state, and hence are proving that the log-negativity is resource independent for all basic unitaries.

We begin with:
\begin{equation}
	K \ket{\Phi_{A}}\ket{\Phi_{B}} = \sum_{k} \lambda_{k} (A_{k} \ket{\Phi_{A}}) \otimes (B_{k} \ket{\Phi_{B}}) , 
\end{equation}
and we shall prove our result by showing that, except for a factor of $\sqrt{d_{a}d_{b}}$ this decomposition is already a Schmidt decomposition.  We proceed by proving a straightforward, and possibly well-known, fact:
\begin{lem}  
\label{LEM:orthonormal}
For orthonormal sets of operators $\{ A_{k} \}$,  the set of states $\{ \ket{\psi_{A_{k}}} = \sqrt{d_{a}} A_{k}\ket{\Phi_{A}} \}$ are orthonormal.
\end{lem}
The lemma is proven in Appendix~\ref{APP:orthonormal}.  From this lemma, it follows that a Schmidt decomposition of the Choi matrix is:
\begin{equation}
	K \ket{\Phi_{A}}\ket{\Phi_{B}} = \sum_{k} \frac{\lambda_{k}}{\sqrt{d_{a}d_{b}}} \ket{ \psi_{A_{K}}} \ket{ \psi_{B_{k}}}. 
\end{equation}
It is well known that the log-negativity of a pure state follows directly from its Schmidt coefficients, such that:
\begin{eqnarray} 
\label{eqn:Choi_entanglement}
  E_{LN}(K \Phi K^{\dagger}) & = & 2 \log_{2} \left(  \sum_{k} \frac{ \lambda_{k}  }{\sqrt{d_{a}d_{b}}} \right) , 
\end{eqnarray}
which holds for any Kraus operator.  For basic unitaries, this achievable entanglement and the upper bound coincide, giving a closed form for the EC, and proving our main result (theorem~\ref{thm:Choi_state}).

We have shown that for a particular class of unitary operations, which includes all two-qubit gates, the maximum increase of log-negativity from one application of the unitary has a simple closed form.  Furthermore, the same choice of initial state produces this maximum.  As such, if we are promised that a unitary belongs to the class of basic unitaries, then no further analysis is required to find the optimal strategy for generating entanglement.   We also have the surprising result that unlike entropic quantities, such as von Neumann entropy, the log-negativty EC is resource independent for 2-qubit unitaries.

Although the Choi matrix entanglement equals the log-negativity EC, this does not entail the same for the EC based on different monotones.  Indeed, Kraus and Cirac~\cite{Kraus01} showed that preparing the Choi matrix does not always maximize the linear-entropy (assuming pure states only).  The counter-example they gave employed unitaries of the form $U=\exp^{i \theta W}$, where $W$ is the two-qubit swap gate, and the Choi matrix was non-optimal when $\theta$ was below a critical value.  In these cases, there is no unique answer to what the best strategy is, as both entanglement monotones are meaningful.  However, the strength of our result is that it provides a nice analytic solution, whereas no general solution for the linear-entropy EC is yet known.

\section{Numerical evidence}
\label{sec:Numerical}

\begin{figure}
\includegraphics{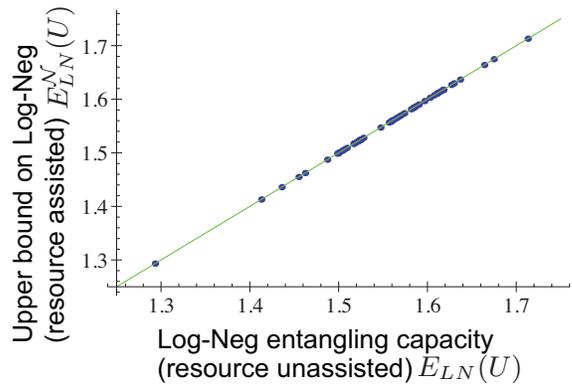}
\caption{Numerically calculated log-negativity EC for randomly selected two-qutrit gates, allowing for two-qutrit ancillary qubits.  The achievable log-negativity with separable states (resource unassisted), $E_{LN} (U)$, is compared to $E^{\mathcal{N}}_{LN} (U)$, which is an upper bound on the maximum resource unassisted increase $E^{\mathcal{D}}_{LN} (U)$.  For all examined instances these quantities are equal (up-to numerical precision of $10^{-6}$).  }
\label{fig:Numerical}
\end{figure}

What do the results so far tell us about the existence of resource independent EC monotones? To recap, we know that log-robustness EC is always resource independent, and that log-negativity EC is resource independent for basic unitaries, such as two-qubit gates.  The most obvious open question is whether log-negativity EC is resource independent for all unitaries, or indeed all CPTP maps, or whether this is a singular feature of 2-qubit gates.  Here we report numerical evidence in favor of the conjecture that for all unitaries we have $E_{LN}^{\mathcal{D}}(U)  = E_{LN} (U)$.

We consider 50 different 2-\textit{qutrit} unitaries randomly selected from the Haar measure, which are typically not basic unitaries.  To calculate $E_{LN} (U)$ we performed a Monte-Carlo search over the set of pure separable states.  We can ignore mixed states because the maximum for any convex function over a convex set is attained by an extremal member of the convex set.  We calculated an upper bound of the resource-assisted EC by considering the dual problem of maximizing over zero-entanglement operators.  Again this optimization can be performed over just the extremal members of the zero-entanglement set $\mathcal{N}$.  Alternatively we could have directly calculated the resource assisted entangling capacity, $E^{\mathcal{D}}_{LN} (U)$, but this would have required~\footnote{This is required because this quantity is neither convex nor a logartihmic function of convex function.} searching over a larger space including non-extremal points.  Surprisingly, this more comprehensive search was not necessary as the upper-bound proved to be tight in all observed instances, as can be seen in figure~\ref{fig:Numerical}.

We allowed for ancilla qubits equal in dimension to the target qubits, and so considered 4-qutrits in total.  Since local maxima are not guaranteed to be global maxima, and the dimensionality ($3^{4}=81$) involved in the problem is quite large, a fairly intensive search must be performed to encourage confidence in having found the global maxima.   Numerical accuracy was benchmarked against unitaries randomly selected from a family of basic unitaries, for which the results always matched our analytic result up-to our working numerical precision of $~10^{-6}$.  Similarly for non-basic unitaries, all instances of numerically found values for $E^{\mathcal{N}}_{LN} (U)$ and $E_{LN} (U)$ differed by no more than this level of precision.

Consequently, it is extremely likely that all two-qutrit unitaries are resource independent.  In other aspects of entanglement theory where special properties have held for two-qubits, but not for higher dimensions, the emergent behavior can be observed for just two-qutrits.  For example, bound entanglement~\cite{HoroBound} and in-equivalence of entangling and disentangling capacity~\cite{DisEnt_Noah} both manifest themselves for qutrit systems and larger.  As such there is little reason to suspect that two-qutrit gates are exceptional in any way, and we are inclined to conjecture that resource independence will hold true for all two-qudit unitaries. 

\section{Tractable Qudit Examples}
\label{examples}

We have seen that various analytic bounds on EC are saturated for two-qubit unitaries, and that the duality upper bound $E^{\mathcal{N}}_{LN} (U)$ seems to be tight in higher dimensions.  In this section we examine some analytically tractable two-qudit unitaries, and show that  in higher dimensions (\textit{i}) preparing the Choi matrix is not always optimal; and (\textit{ii}) the upper-bound expressed by theorem~\ref{eqn:core_result} is not always saturated.  There are many ways to generalize two-qubit control-not gates, with some generalizations giving basic unitaries, but we shall consider the non-basic (for $d>2$) family that acts on systems $a$ and $b$ of dimension $d$:
\begin{equation}
	U_{d} = ( \unity - \kb{0}{0} )  \otimes \unity + \kb{0}{0} \otimes X_{d} ,
\end{equation}
where $X_{d}$ is the generalized Pauli bit-flip operator:
\begin{equation}
	X_{d}  =  \sum_{j=0}^{d-1} \kb{j \oplus 1}{j},
\end{equation}
with $\oplus$ indicating addition modulo $d$. A slight rearrangement gives the Schmidt decomposition with correctly normalized operators:
\begin{equation*}
	U_{d} =  \sqrt{d(d-1)}  \left(  \frac{\unity - \kb{0}{0}}{\sqrt{(d-1)}} \right)  \otimes \frac{\unity}{\sqrt{d}} + \sqrt{d}  \kb{0}{0} \otimes \frac{X_{d}}{\sqrt{d}} .
\end{equation*}
Given that the Schmidt coefficients are $\sqrt{d(d-1)}$ and $\sqrt{d}$, the entanglement of the Choi matrix is simply (see Eqn.~\ref{eqn:Choi_entanglement}):
\begin{eqnarray}
	E_{LN}(U_{d} \Phi U_{d}^{\dagger}) & = & 2 \log_{2} \left(  (\sqrt{d(d-1)}+\sqrt{d})/d  \right) , \\ \nonumber
 & = & 2 \log_{2} \left(  \sqrt{1-d^{-1}}+ \sqrt{d^{-1}}  \right) ,
\end{eqnarray}
which vanishes in the limit of large $d$.  

However, if we consider the input state $\ket{\psi}=(\ket{0}+\ket{1})\ket{0}/\sqrt{2} $, then it is easy to check that $U_{d}\ket{\psi}$ produces one e-bit of entanglement, or $\log_{2}(2)$ of log-negativity.  Consequently, for $d>2$, preparing the Choi matrix is not optimal.  No operator with only two non-zero Schmidt coefficients can produce more than $\log_{2}(2)$ log-negativity and so this must be the entangling capacity for all values of $d$.

We now consider the value of the upper-bound imposed by theorem~\ref{eqn:core_result}, which tells us that:
\begin{eqnarray}
  E_{LN}(U_{d}) & \leq & \log_{2} \left( || O_{A} ||_{\mathrm{op}} . || O_{B} ||_{\mathrm{op}} \right) ,\\ \nonumber
  & = &\log_{2}\left(   \left( \sqrt{d} \right) .\left(\sqrt{1-d^{-1}}+ \sqrt{d^{-1}}  \right) \right), \\ \nonumber
  & = & \log_{2}\left( \sqrt{d-1}+ 1  \right),
\end{eqnarray}
which goes to infinity as $d$ increases, and hence this is also not tight against the actual EC of $\log_{2}(2)$.

On a more positive note, we observe that these unitaries are also resource independent.  Again this provides further support for our conjecture. 

\section{conclusions}

We have investigated the maximum entanglement that can be achieved by a dynamic process allowing for use of ancillary Hilbert space, with particular attention paid to the boost that can sometimes be achieved by exploiting prior entanglement.  These quantities depend on the underlying entanglement monotone that is used to quantify the entanglement of the quantum state.  We considered a family of monotones that includes log-negativity and log-robustness, with a particular focus on log-negativity.  We found that this family of entangling capacity monotones is often insensitive to the availability of prior entanglement, a property we call resource independence of the entangling capacity.  Our most general result, which covers a whole family of entangling capacity monotones, we call the duality theorem as it tells us that the resource-assisted entangling capacity cannot exceed the solution of a dual problem determined by the structure of the entanglement monotone.  

The duality theorem immediately entails that log-robustness is resource independent for all dynamical operations.  For log-negativity, the duality theorem entails a weaker result with a murkier interpretation.  However, further results built on the duality theorem prove that log-negativity entangling capacity is resource independent for all two-qubit unitaries, and other so-called basic unitaries.  Furthermore, we presented numerical evidence that two-qutrit unitaries are resource independent. This leads us to the conjecture that all unitary gates are resource independent for this particular metric of entangling capacity.   Settling this conjecture is an especially interesting open problem.

We have also shown that the log-negativity entangling capacity of two-qubit gates has a closed form, and equals the entanglement of the corresponding Choi matrix.  For no other measure of entangling capacity has a closed form been found, and so this result may prove extremely useful in simplifying the analysis of entangling capacity.   We have seen that beyond two-qubit unitaries the Choi matrix no longer captures the log-negativity entangling capacity.  However, the simplicity of the result for two-qubit unitaries is encouraging evidence that a closed form may be also be found for higher dimensional unitaries, and maybe even general dynamical maps.
 
\section{Acknowledgements}

We would like to thank Akihito Soeda, Dan Browne, Hussain Anwar, Matty Hoban, and Joe Fitzsimons, for interesting discussions on this topic and useful comments on the paper.   We would also like to thank Daniel Burgarth for conversations about the Choi-Jamiolkowski isomorphism, which inspired the early stages of this work.  This research was supported by the Royal Commission for the Exhibition of 1851.

\appendix

\section{}
\label{APP:subadd_necessary}

In the papers preamble we asserted that subadditivity of an entanglement monotone is a necessary property for the corresponding EC to be resource independent.  The basic intuition is that when subadditivity is violated, one can boost entangling capacity by simply appending an ancillary entangled state that actually plays no active role.   Indeed, this spurious boost may ensure an infinite amount of resource assisted EC for almost all CPTP maps, making the quantity devoid of concrete meaning.  This rough intuition will suffice for understanding our main results, but for the interested reader we shall provide a rigorous proof here.

If an entanglement monotone is not subadditive, then there must exist at least two quantum states $\rho_{1}$ and $\rho_{2}$, such that:
\begin{equation}
	E (\rho_{1} \otimes \rho_{2}) > E (\rho_{1})+ E (\rho_{2}) .
\end{equation}
We now define a CPTP map that will violate resource independence:
\begin{equation}
	\Lambda_{1}( \sigma ) = \tr_{a,b} (   \kb{0,0}{0,0} \sigma ) \rho_{1} +  (1- \tr_{a,b} (   \kb{0,0}{0,0} \sigma )) \unity / d,
\end{equation}
where $d$ is the dimension of $\rho_{1}$, and $\tr_{a,b}(...)$ is a partial trace over target qubits $a$ and $b$. Clearly, with the initially separable state $\ket{0,0}$ and one application of $\Lambda_{1}$ we can produce the output $\rho_{1}$.  For any other separable input, $\sigma_{A} \otimes \sigma_{B}$, we have that:
\begin{equation}
	\Lambda_{1}( \sigma_{A} \otimes \sigma_{B} )	= \eta \rho_{1} \otimes \sigma_{a'} \otimes \sigma_{b'} + (1-\eta)  \frac{\unity}{d} \otimes \sigma'_{a'} \otimes \sigma'_{b'},
\end{equation}
where:
\begin{eqnarray*}
	\eta & = & \tr (  \kb{0,0}{0,0} \sigma ) , \\
	\sigma_{a'} \otimes \sigma_{b'} & = & \tr_{a,b}  (   \kb{0,0}{0,0} \sigma ) / \eta , \\
	\sigma_{a'} \otimes \sigma_{b'} & = & \tr_{a,b}  (  (\unity - \kb{0,0}{0,0}) \sigma ) / (1-\eta) .
\end{eqnarray*}
The above output state can always be produced from $\rho_{1}$ by local operations and classical communication. Therefore, since $E$ is monotonic under such operations the resource unassisted EC cannot exceed $E(\rho_{1})$, and we have $E(\Lambda_{1})=E(\rho_{1})$.

We will now prove that the resource assisted EC can always exceed this value.  We consider the entangled input state $\kb{0,0}{0,0}\otimes \rho_{2}$ where $\rho_{2}$ is in the ancillary Hilbert space $H_{a'} \otimes H_{b'}$.  Since the ancillary qubits are uncorrelated w.r.t the target qubits, this ancillary resource plays no active role in the dynamical process, such that:
\begin{equation}
	\Lambda_{1}( \kb{0,0}{0,0} )  \otimes \rho_{2} = \rho_{1} \otimes \rho_{2},
\end{equation}
the entanglement of which is:
\begin{eqnarray}
	E ( \Lambda_{1} ( \kb{0,0}{0,0} )  \otimes \rho_{2} )  & = & E (\rho_{1} \otimes \rho_{2} ), \\ \nonumber
	& > &  E (\rho_{1}) + E( \rho_{2} ),
\end{eqnarray}
where the second line uses the violation of subadditivity.  It follows directly that $E^{\mathcal{D}}(\Lambda_{1})>E(\Lambda_{1})$, and so the EC is not resource independence.  
  
\section{}
\label{APP:always_break_subadd}

We show here that an entanglement monotone, $E_{1}$, that is subadditive, can always be rescaled into a non-subadditive monotone.  Consequently, a resource independent EC can always lose this property by appropriate rescaling.  A second monotone $E_{2}$ is a rescaling of $E_{1}$ if and only if there exists a monotonically increasing function, $g$, such that $E_{2}(\rho)=g(E_{1}(\rho))$.  We proceed under the weak assumption that if $E_{1}(\rho)>0$, then
\begin{equation}
	E_{1} (\rho \otimes \rho ) >  E_{1}(\rho) .
\end{equation}
This assumption is satisfied for all well-known entanglement monotones, and only fails to hold for extremely contrived constructions.  We can now break subadditivity for $\rho \otimes \rho$, by choosing a rescaling function such that:
\begin{equation}
	g  ( x + \delta ) > 2 g (x),
\end{equation}
where $x=E_{1}(\rho )$, and $x + \delta =E_{1}(\rho \otimes \rho)$.  Our earlier assumption ensures that $\delta > 0$, and so we simply choose a rescaling function that more than doubles in the gap between $x$ and $x+\delta$.

Since a rescaling can always break resource independence, it is reasonable to ask whether the inverse operation can always be performed.  That is, can we always rescale a monotone to create a resource independent EC? We do not currently have a solution to this problem.  However, one fact is clear: rescaling a strongly-additive resource-dependent entanglement monotone can not make it resource independent without also breaking strong additivity. 

\section{}
\label{APP:duality_details}

In this appendix we compare the entanglement of $\Lambda( \rho^{*})$ with that of $\Lambda( \sigma^{\pm})$, which results in the inequality in Eqn.~\ref{EQN:duality_half_step}. Recall that the zero-entanglement operators, $\sigma^{\pm}$, are given by the decomposition of $\rho^{*}$ in Eqn.~\ref{EQN:opt_rho_decomp}.  

We begin by considering $\Lambda(\sigma^{\pm})$, where the CPTP map, $\Lambda$, preserves the trace and Hermiticity of $\sigma^{\pm}$, and hence the new operators also have optimal decompositions into zero-entanglement operators, such that:
\begin{eqnarray}
	\Lambda( \sigma^{+} ) & = & (1+ a) \sigma^{+, +}  - a \sigma^{+, -} , \\
	\Lambda( \sigma^{-} ) & = & (1+ b) \sigma^{-, +}  - b \sigma^{-, -} .
\end{eqnarray}
Combining these equations with Eqn.~\ref{EQN:Lambda_rho_decomp}, we have that:
\begin{eqnarray}
	\Lambda (\rho^{*})  & = &  (1+t)(1+a) \sigma ^{+,+} - (1+t)a \sigma ^{+,-} ,  \\ \nonumber
	& & - t (1+b) \sigma ^{-,+} + t b \sigma ^{-,-}.
\end{eqnarray}
We can simplify this expression by defining new operators
\begin{eqnarray}
	\tilde{\sigma}^{1} & = & \frac{(1+t)(1+a) \sigma ^{+,+} +  t   b \sigma ^{-,-}}{ (1+t)(1+a)  + t b } , \\ \nonumber
	\tilde{\sigma}^{2} & = & \frac{ (1+t)a  \sigma ^{+,-} + t (1+b) \sigma ^{-,+}}{ (1+t)a +  t (1+b)   } .
\end{eqnarray}
Since the zero-entanglement operators form a convex set, it follows that $\tilde{\sigma}^{1,2}$ are also zero-entanglement operators.  In terms of these new operators we have:
\begin{eqnarray}
	\Lambda (\rho^{*})  & = & (1+t') \tilde{\sigma}^{1} \nonumber - t' \tilde{\sigma}^{2}	 ,
\end{eqnarray}
where,
\begin{equation}
	t' = (1+t)a +  t (1+b).
\end{equation}
This decomposition of $\Lambda (\rho^{*})$ is of the correct form for calculating the entanglement, but may not be the decomposition that minimizes $t'$.  Hence, it gives an upper bound on the entanglement:
\begin{eqnarray*}
	E_{\mathcal{M}}(  \Lambda(\rho^{*})  )  & \leq & t'  = ((1+t)a +  t (1+b)  )   	.
\end{eqnarray*}
Next, we define $c=\max \{ a, b \}$, and use it to replace these variables.  This replacement further increases the larger side of the inequality, and so we have that:
\begin{eqnarray}
	E_{\mathcal{M}}( \Lambda( \rho^{*})  )   & \leq & (1+t)c+t(1+c)  .
\end{eqnarray}
Adding contributions to both sides gives a useful factorization
\begin{eqnarray}
1+2 E_{\mathcal{M}}( \Lambda( \rho^{*})  )  & \leq &   (1 + 2c)(1+2t)    ,  \\ \nonumber
& = & (1 + 2c)(1+2 E_{\mathcal{M}}( \rho^{*} )).
\end{eqnarray}
Taking the logarithm, and applying Eqn~(\ref{eqn:Log}), gives:
\begin{eqnarray}
	E_{L\mathcal{M}}( \Lambda(\rho^{*}) ) \leq \log_{2} (1 + 2c) + E_{L \mathcal{M}}(\rho^{*}) .
\end{eqnarray}
Subtracting the initial entanglement gives:
\begin{equation}
 E_{L \mathcal{M}}( \Lambda (\rho^{*}) ) - E_{L \mathcal{M}}(\rho^{*}) \leq \log_{2} (1+ 2c) .
\end{equation}
Notice how the terms involving the variable $t$ have cancelled out, simplifying the nature of the bound.  Next, we observe that the following three facts (\textit{i}) $c=\max \{ a,b \}$; (\textit{ii}) $1+2a=E_{L\mathcal{M}}(\Lambda( \sigma^{+}))$; and (\textit{iii}) $1+2b=E_{L\mathcal{M}}(\Lambda( \sigma^{-}))$; joinly entail that:
\begin{equation}
	1+2c = \max \{ E_{L\mathcal{M}}(\Lambda( \sigma^{\pm})) \} ,
\end{equation}
and hence:
\begin{equation}
 E_{L \mathcal{M}}( \Lambda (\rho^{*}) ) - E_{L \mathcal{M}}(\rho^{*}) \leq \max \{ E_{L\mathcal{M}}(\Lambda( \sigma^{\pm}))  \} .
\end{equation}
This complete derivation of the inequality.

\section{}
\label{APP:basic_unitaries}

In general, to search for a basic decomposition of a given $U$ we find an orthonormal Schmidt decomposition.  This will be a basic decomposition if each term is proportional to a unitary.  If all the non-zero Schmidt coefficients are different, then there is a unique Schmidt decomposition of $U$, and finding it gives a conclusive answer as to whether $U$ is basic.  Whereas degeneracy in the Schmidt coefficients entails a family of valid Schmidt decompositions and the problem is more involved; in these instances one must exhaustively search the family of Schmidt decompositions for one that is basic. 

However two-qubit unitaries are always basic and we shall give a proof here.  This feature of two-qubit unitaries follows from the widely exploited~\cite{Kraus01,Leifer03,Nielsen03} fact that all two-qubit unitaires are local unitary equivalent to a \textit{diagonal} operator $U_{\mathrm{diag}}$, such that:
\begin{equation}
	U = ( U_{A} \otimes U_{B} ) U_{\mathrm{diag}}   (V_{A} \otimes V_{B} ) ,
\end{equation}
where the diagonal operator has the form:
\begin{eqnarray}
	U_{\mathrm{diag}}  & = & \exp^{i ( \sum_{j=1}^{3} c_{j} \sigma_{j} \otimes \sigma_{j} )}  ,
\end{eqnarray}
where $\sigma_{1,2,3}$ are the Pauli spin matrices and the operator is unitary for all real $c_{j}$.  Expanding out the matrix exponential, and including the local unitaries we have:
\begin{eqnarray}
     	U = ( U_{A} \otimes U_{B} ) (  \sum_{j=0}^{4} a_{j} \sigma_{j} \otimes \sigma_{j} )   (V_{A} \otimes V_{B} ) ,
\end{eqnarray}
where $a_{j}$ are complex numbers determined by $c_{j}$.  We define $A_{j}=U_{A} \sigma_{j} V_{A} / \sqrt{d_{a}}$ and $B_{j}= \mathrm{Arg}( a_{j} )  U_{B} \sigma_{j}V_{B}/ \sqrt{d_{b}}$, with any complex phase absorbed by the operators.  It is straightforward to check that $\{ A_{j} \}$ and $\{ B_{j} \}$  are orthonormal sets and are proportional to unitaries.  Consequently, all two-qubit unitaries have a basic decomposition with $\lambda_{j}=|a_{j}|$. 

\section{}
\label{APP:Log_Neg_Bound}

This appendix takes equation~(\ref{EQN:Log_neg_partial_trace}), and derives an upper bound on its trace norm, producing equation~(\ref{EQN:Log_neg_trace_norm}).  As outlined in the main text, the overall strategy is to find a valid, but not necessarily minimal, decomposition.  We begin by introducing a more compact notation:
\begin{eqnarray}
	(K \sigma K^{\dagger})^{\Gamma} & = &  \sum_{k,j} \lambda_{k} \lambda_{j} M_{j,k}  ,
\end{eqnarray}
where,
\begin{equation}
	M_{j,k} =  (A_{k}  \otimes B_{j}^{T}) \sigma^{\Gamma}  (A_{j} \otimes  B_{k}^{T})^{\dagger} .
\end{equation}
The trace norm is evaluted by searching over real decompositions of positive unit-trace operators.    Recall that for any positive Hermitian matrix, $\rho$, the matrix $C \rho C^{\dagger}$ is also positive and Hermitian for any $C$.  Consequently, the diagonal terms $M_{k,k}$ are always positive and Hermitian.  However, the cross terms $M_{j,k \neq j}$ are not individually positive, and so the trick is to find an appropriate decomposition for collections of cross terms.  It turns out that we can find such a decomposition for the sum of paired cross terms, $M_{j,k}+M_{k,j}$, such that:
\begin{eqnarray}
	M_{j,k} + M_{k,j} = \tilde{M}_{j,k}^{+} - \tilde{M}_{j,k}^{-} ,
\end{eqnarray}
where the new operators are:
\begin{eqnarray}
	\tilde{M}_{j,k}^{ \pm } & = & \frac{1}{2} (X_{j,k} \pm X_{k,j}) \sigma^{\Gamma} (X_{j,k} \pm X_{k,j})^{\dagger}  ,
\end{eqnarray}
and the $X_{j,k}$ are operators:
\begin{eqnarray}
	X_{j,k} & = &  A_{k} \otimes B_{j}^{T}  .
\end{eqnarray}
Now our tilded operators, $\tilde{M}_{j,k}^{\pm}$, \textit{are} positive matrices.  Collecting these equations together, we have that:
\begin{eqnarray*} 
	(K \sigma K^{\dagger})^{\Gamma} & = &  \sum_{k}  \lambda_{k} ^{2} M_{k,k} + \sum_{k,j<k} \lambda_{k}\lambda_{j}( \tilde{M}_{j,k}^{+}  - \tilde{M}_{j,k}^{-} ) ,
\end{eqnarray*}
where each of the operators is positive and Hermitian. Now we can use our decomposition, which is not necessarily optimal, to prove a bound on the negativity, such that:
\begin{eqnarray*} 
	||(K \sigma K^{\dagger})^{\Gamma}|| & \leq &  \sum_{k}  \lambda_{k} ^{2} \tr( M_{k,k} ) \\ \nonumber & & + \sum_{k,j<k} \lambda_{k}\lambda_{j}( \tr ( \tilde{M}_{j,k}^{+} )  + \tr( \tilde{M}_{j,k}^{-} )) .
\end{eqnarray*}
Again we can further simplify pairs of terms:
\begin{eqnarray}
	\tr (  \tilde{M}_{j,k}^{+}  +  \tilde{M}_{j,k}^{-}  ) & = & \tr( X_{j,k} \sigma^{\Gamma} X_{j,k}^{\dagger} + X_{k,j} \sigma^{\Gamma} X_{k,j}^{\dagger} )  , \\ \nonumber
	& = & \tr( X_{j,k}^{\dagger} X_{j,k} \sigma^{\Gamma} ) +\tr (X_{k,j}^{\dagger}  X_{k,j} \sigma^{\Gamma} )  , 
\end{eqnarray}
where the second line follows from the linearity and cyclicality of the trace.  Since the diagonal terms similarly satisfy $\tr( M_{k,k} )=\tr( X_{k,k}^{\dagger}X_{k,k} \sigma^{\Gamma} )$, we have that:
\begin{eqnarray*} 
	||(K \sigma K^{\dagger})^{\Gamma}|| & \leq &  \sum_{j, k}  \lambda_{k}  \lambda_{j} \tr ( X_{j,k}^{\dagger} X_{j,k} \sigma^{\Gamma} ) , \\
		 & = &  \sum_{j, k}  \lambda_{k}  \lambda_{j} \tr ( A_{j}^{\dagger}A_{j} \otimes B_{k}^{*}B_{k}^{T} \sigma^{\Gamma} ) . 
\end{eqnarray*}
This completes our proof.

\section{}
\label{APP:orthonormal}
Here we prove lemma.~\ref{LEM:orthonormal}, as follows:
\begin{eqnarray}
 \bk{\psi_{A_{k}}}{\psi_{A_{j}}  } & = & d_{a} \bra{\Phi_{A}} (A_{k}^{\dagger} \otimes 1)(A_{j} \otimes 1)  \ket{\Phi_{A}  }  \\ \nonumber
 & = &  \sum_{x,y} \bra{x,x}  (A_{k}^{\dagger} \otimes 1)(A_{j} \otimes 1)  \ket{y,y}  \\ \nonumber
  & = &  \sum_{x,y} \bra{x}  A_{k}^{\dagger} A_{j}  \ket{y} \bk{x}{y}   \\ \nonumber
    & = &  \sum_{x,y} \bra{x}  A_{k}^{\dagger} A_{j}  \ket{y} \delta_{x,y}  \\ \nonumber
 & = &  \sum_{x} \bra{x}  A_{k}^{\dagger} A_{j}  \ket{x}   .
\end{eqnarray} 
The final line is simply the trace operation performed in the computational basis, and so:
\begin{eqnarray}
 \bk{\psi_{A_{k}}}{\psi_{A_{j}}  }  & = &  \tr \left( A_{k}^{\dagger} A_{j} \right)    \\ \nonumber
    & = &  \delta_{k,j}   
\end{eqnarray}
where we have used that the operators $\{ A_{j} \}$ are orthonormal.  A similar proof holds for system $B$.

\end{document}